\documentclass[11pt, a4paper]{article}
\usepackage[a4paper, margin=2.5cm]{geometry}
\usepackage[utf8]{inputenc}
\usepackage[T1]{fontenc}

\usepackage{amsmath, amssymb}
\usepackage{booktabs}
\usepackage[authoryear, round]{natbib}
\usepackage[colorlinks=true, linkcolor=blue, citecolor=blue, urlcolor=blue]{hyperref}
\usepackage{graphicx}

\usepackage{orcidlink}
\renewcommand{\orcidlink}[1]{\href{https://orcid.org/#1}{\includegraphics[width=10pt]{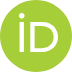}}} 

\newcommand{\Bspot}{B_\mathrm{spot}}
\newcommand{\DFspot}{\Delta F_\mathrm{spot}}
\newcommand{\lDspot}{\log(\DFspot)}

\title{\textbf{Unraveling the magnetic fields of starspots with transit mapping}}
\author{
\textbf{F. Menezes}$^1$ \orcidlink{0000-0002-4703-4027},
A. Valio$^1$ \orcidlink{0000-0002-2106-4332}, 
\& A. Araújo$^1$ \orcidlink{0000-0002-1671-8370}
}

\date{\small $^1$ Centro de R\'adio Astronomia e Astrof\'isica Mackenzie (CRAAM), Universidade Presbiteriana Mackenzie, Rua da Consola\c{c}\~ao, 930, S\~ao Paulo, SP, 01302-907, Brazil}

\begin{document}

\maketitle

{\noindent Contact: \url{menezes.astroph@gmail.com}}\\
{\noindent Published in \textit{Solar and Stellar Multi-scale Activity, Proceedings IAU Symposium No. 400, 2026}, S. Vargas-Domínguez \& J. C. Buitrago-Casas, eds.} \\
{\noindent Available at \url{https://doi.org/10.1017/S174392132610725X}}

\begin{abstract}
Stellar magnetic activity is known to decrease as stars age, but the underlying physical mechanism for this decline remains a key question. Does this evolution occur because individual starspots become magnetically weaker (a change in \textit{quality}), or because the total area covered by spots diminishes (a change in \textit{quantity})? We address this question by synthesizing results from two recent studies that leverage starspot transit mapping of Kepler and CoRoT targets. First, we show that the magnetic field strength of starspots remains consistently high, on a \textit{saturation plateau} of approximately 3500 G, regardless of stellar age or rotation period. This suggests that the intrinsic quality of spots does not significantly evolve. Second, we demonstrate a strong anticorrelation between starspot area coverage and stellar age, with young stars exhibiting large filling factors (up to 29\%) that decrease sharply over time. Our combined results provide compelling evidence that the evolution of stellar magnetic activity is primarily driven by a change in the quantity of magnetic flux emerging on the surface, not by a change in the intrinsic strength of the spots themselves. These findings have direct implications for refining stellar dynamo models and improving age-dating techniques.
\end{abstract}


\maketitle

\section{Introduction}

Starspots, analogous to sunspots, are the most prominent manifestations of stellar magnetic activity. Their properties provide crucial insights into the stellar dynamo, the mechanism responsible for generating magnetic fields in cool stars \citep{Berdyugina2005}. A deeper understanding of the analogy between solar and stellar spots is crucial for stellar dynamo comprehension, an effort supported by advances in solar and stellar observations \citep{Wedemeyer2024}. It is a well-established empirical fact that a star's magnetic activity declines as it ages, a cornerstone of age-dating techniques like gyrochronology \citep{Skumanich1972, Barnes2007}. However, the physical process driving this decline is not fully understood, and recent studies have begun to explore not only internal mechanisms but also external influences, such as star-planet interactions \citep{Menezes2025}.

The manifestation of stellar magnetic activity includes phenomena such as chromospheric emission, coronal X-rays, flares, and most visibly, starspots. Measuring the magnetic fields inside these spots is a difficult task. Traditional methods face limitations; for instance, as discussed by \citet{Donati2009}, Zeeman-Doppler Imaging (ZDI) is best suited for mapping global, weaker magnetic fields on fast-rotating stars and often misses the compact, ultra-strong fields within spots. Zeeman Broadening (ZB) can measure total magnetic flux but is limited to slowly rotating stars and is entangled with the spot filling factor \citep{Reiners2012}.

The fundamental question of this work is whether the age-related activity decrease is driven by a change in the intrinsic properties of starspots or by their surface coverage. As a star ages, do its spots become magnetically weaker (a change in \textit{quality}), or does the star simply produce fewer spots, thus reducing the total spotted area (a change in \textit{quantity})? This work synthesizes the key findings from two recent studies \citep{Menezes2024, Araujo2025} that applied the starspot transit mapping technique, first proposed by \citet{Silva2003}, to address the quality vs. quantity dichotomy.

\section{Methodology}

\begin{figure}[h!]
\centering
\includegraphics[width=0.65\linewidth]{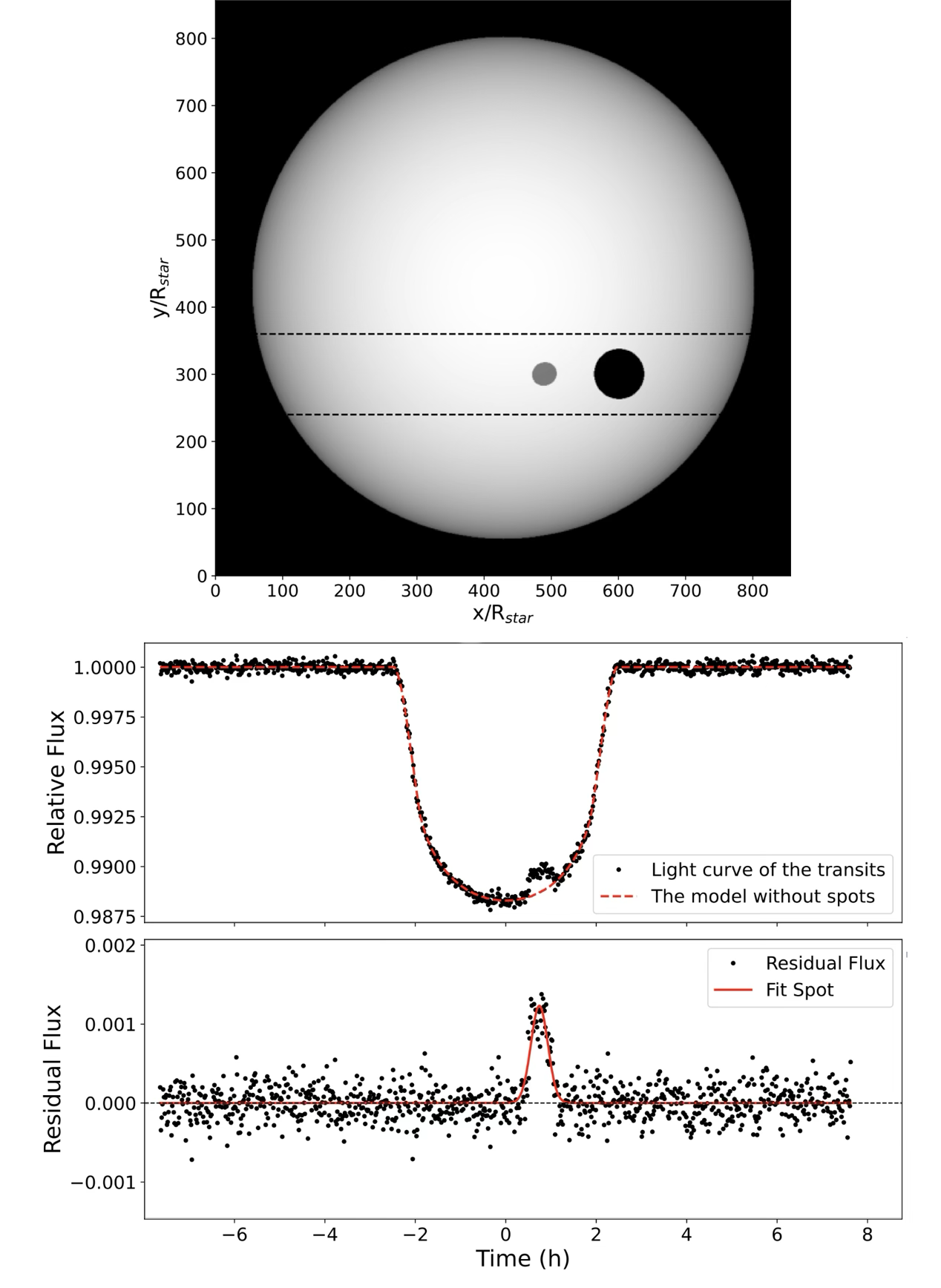}
\caption{Illustration of the starspot transit mapping technique. 
\textit{Top panel:} A schematic of a transiting exoplanet (large black circle) occulting a starspot (smaller grey circle) on the stellar disk. 
\textit{Middle panel:} The observed transit light curve (black dots) shows a characteristic anomaly, or bump, where the planet crosses the spot. A model of a transit on an immaculate star is overplotted (red dashed line). 
\textit{Bottom panel:} The residual flux, obtained by subtracting the spotless model from the data. The spot-crossing event is isolated as a prominent peak, which is modeled (red solid line) to determine the spot's physical properties.}
\label{fig:transit}
\end{figure}

Our approach leverages the unique geometry of planetary transits to probe the properties of starspots. When a planet passes in front of its host star, it causes a characteristic dip in the observed stellar brightness. If the planet's path crosses a dark starspot during transit, the amount of blocked light temporarily decreases, creating a small positive "bump" in the transit light curve (Figure~\ref{fig:transit}).

By carefully analysing these spot-crossing events, we can extract key physical parameters of the starspot, including its size, position, and intensity contrast relative to the surrounding photosphere. This technique offers several advantages:
\begin{itemize}
\item It is not limited by stellar rotation rate, unlike Zeeman-Doppler Imaging.
\item It provides direct measurements of individual spots, avoiding the degeneracy between spot coverage and field strength that affects Zeeman Broadening.
\item It can be applied to archival data from space-based photometric missions like Kepler, K2, and TESS.
\end{itemize}

\subsection{Deriving Magnetic Field Strength and Area Coverage}
A key challenge in modelling spot-crossing events is the well-known degeneracy between a spot's area (A) and intensity (I). To break this degeneracy, we use a combined parameter, the Spot Flux Deficit ($\DFspot$), which robustly quantifies the total missing flux from the spot:
\begin{equation}
\DFspot = A \times (1 - I)
\end{equation}
As shown in \citet{Menezes2024}, the strong empirical relationship between the Spot Flux Deficit and the maximum magnetic field strength ($\Bspot$), calibrated using over 32,000 sunspots \citep{Valio2020}, is given by:
\begin{equation}
\Bspot = (1170 \pm 3) + (844 \pm 3) \times \lDspot
\end{equation}
This relationship, calibrated using solar observations where both photometric properties and direct magnetic field measurements are available, allows us to estimate the magnetic field strength of starspots on distant stars where direct magnetic measurements are not feasible. The same models also allow us to estimate the fractional area of the star covered by spots, as described in \citet{Araujo2025}.

\subsection{Sample Selection and Data Analysis}

Our study includes 14 FGK and M stars hosting transiting exoplanets from the Kepler, K2, and TESS missions, spanning ages from $\sim$100 Myr to $\sim$4.5 Gyr. For each star, we identified spot-crossing events, measured the spot properties, and derived the corresponding magnetic field strengths and area coverages to investigate trends with stellar age.

\section{Results}

\subsection{Magnetic Field Strengths of Starspots}

\begin{figure}[h!]
\centering
\includegraphics[width=0.55\textwidth]{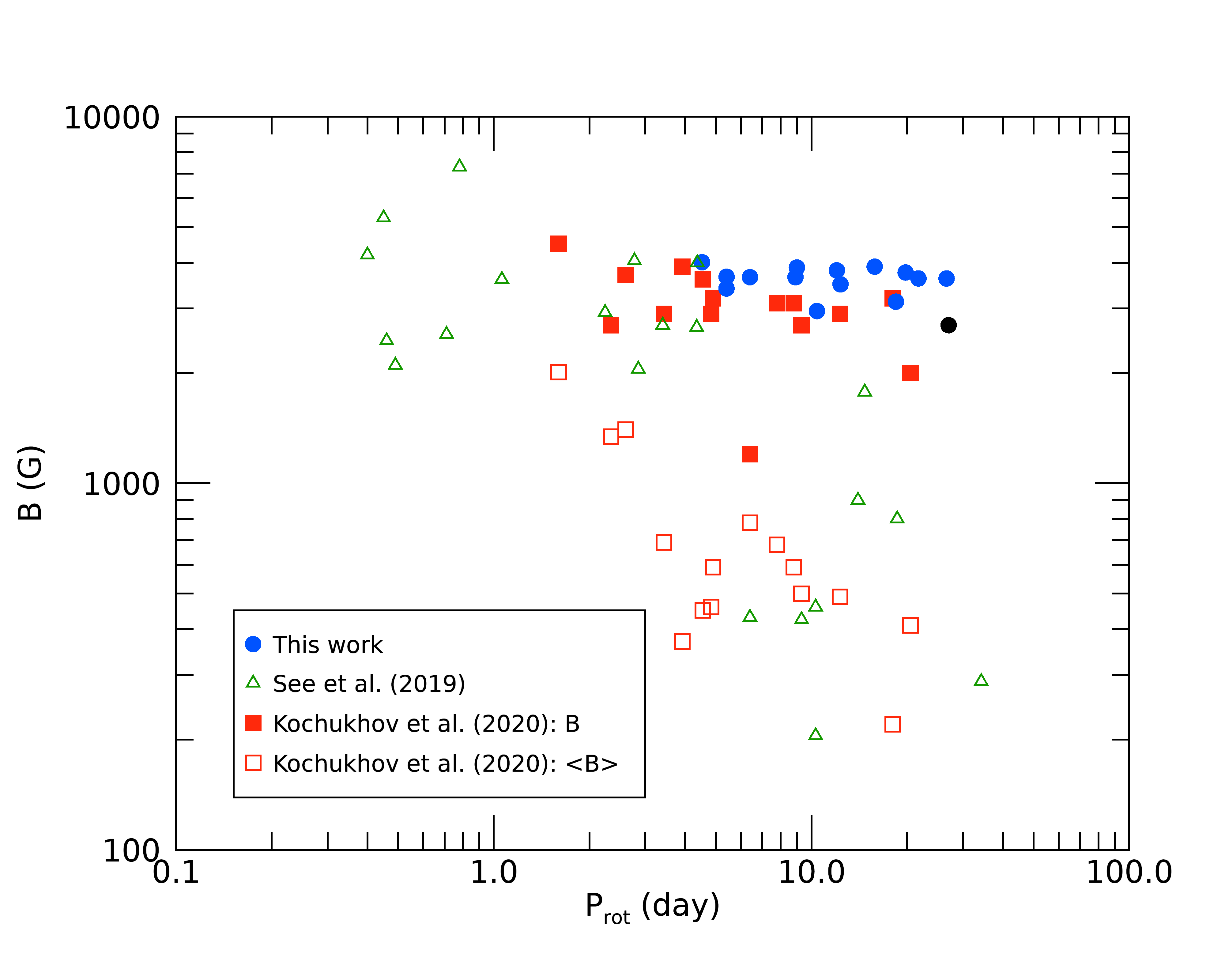}
\caption{The magnetic field strength of starspots (red circles) as a function of stellar rotation period. Our results remain on a high "saturation plateau," in contrast to measurements of the global mean magnetic field (open symbols from other works), which decrease with longer periods. Adapted from \citet{Menezes2024}.}
\label{fig:plateau}
\end{figure}

Our analysis of 2,269 spots on 14 stars revealed that the average spot magnetic fields are consistently high, ranging from 2,700 to 4,600 G, with a sample average of approximately 3500 $\pm$ 300 G \citep{Menezes2024}. Remarkably, we found no significant correlation between spot magnetic field strength and stellar age or rotation period (see Figure~\ref{fig:plateau}). Even on relatively old, slowly rotating stars, the spots maintain field strengths comparable to those on young, rapidly rotating stars. This finding suggests the existence of a magnetic saturation plateau for starspots.

\subsection{Spot Coverage and Stellar Age}

\begin{figure}[h!]
\centering
\includegraphics[width=0.65\textwidth]{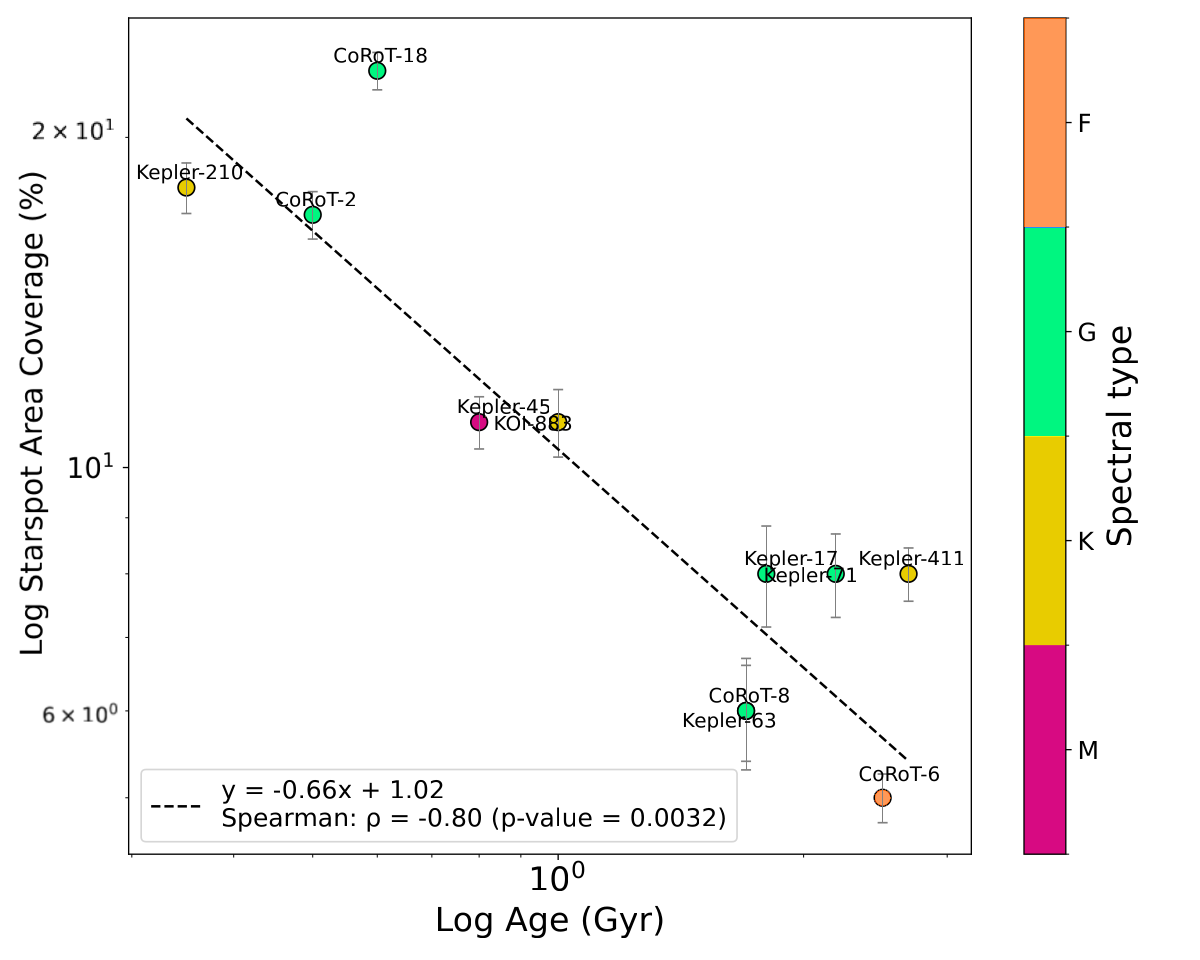}
\caption{The logarithm of starspot area coverage as a function of stellar age. The plot shows a strong anticorrelation, confirming that older stars have a significantly smaller fraction of their surface covered by spots. From \citet{Araujo2025}.}
\label{fig:coverage}
\end{figure}

In contrast to the consistent magnetic field strengths, our analysis of spot coverage on 11 stars revealed a strong and statistically significant anticorrelation with stellar age (Spearman $\rho = -0.80$), as shown in Figure \ref{fig:coverage} \citep{Araujo2025}. Young stars are heavily spotted, with up to 29\% of their surface covered. In comparison, older stars (age $\sim$1-4.5 Gyr) exhibited much lower coverage, typically around 4-10\%. This provides compelling evidence that the quantity, rather than the quality, of starspots is the primary driver of the observed decline in stellar magnetic activity.

\section{Discussion: Quality versus Quantity}

\subsection{The "Quality": Spot Magnetic Field Strength Remains Constant}

Most critically, we found that the spot magnetic field strength does not decrease with stellar age or with longer rotation periods. As shown in Figure \ref{fig:plateau}, our measurements (red circles) lie on a \textit{saturation plateau}, even for slowly rotating stars. The discovery of a magnetic saturation plateau for starspots has profound implications for our understanding of stellar dynamo processes. It suggests that the mechanism generating the intense magnetic fields within spots operates at or near its maximum efficiency across a wide range of stellar conditions. Once a spot forms, the local magnetic field amplification appears to proceed to a similar endpoint regardless of the star's global properties.

This saturation behavior may reflect fundamental physical constraints on the local dynamo process, such as:

\begin{itemize}
\item Equipartition between magnetic and kinetic energy densities in the convection zone.
\item Thresholds for magnetic buoyancy and flux emergence.
\item Limits imposed by magnetohydrodynamic instabilities.
\end{itemize}

The consistency of spot magnetic field strengths across different stellar types and ages also suggests that the basic physics of spot formation remains similar throughout main-sequence evolution, even as the global dynamo efficiency changes.This indicates that the physical process generating the magnetic field inside a spot is highly stable and does not weaken significantly as the star evolves. This finding suggests that the decline in global activity is not caused by a change in the intrinsic \textit{quality} of spots.

\subsection{The "Quantity": Starspot Area Coverage Decreases with Age}

If the strength of spots remains constant, the decline in activity must be driven by a change in their number or area. Our second study \citep{Araujo2025}, tested this by measuring the starspot area coverage on 11 stars of various ages. Our findings provide strong evidence that the well-documented decline in stellar magnetic activity with age is driven primarily by a reduction in the number and total area of starspots, rather than by a weakening of the spots themselves. This provides direct evidence that the \textit{quantity} of spots decreases dramatically as a star ages.

This ``quantity over quality" conclusion has several important implications:

First, it suggests that age-related changes in the stellar dynamo affect primarily the generation and emergence rate of magnetic flux tubes, rather than the intensity of the magnetic fields they contain. This points to large-scale properties of the dynamo, such as differential rotation or meridional circulation, as the key factors that evolve with stellar age.

Second, it helps explain why activity indicators that measure the integrated effect of magnetic regions (such as chromospheric emission or X-ray luminosity) show smooth declines with age, while the properties of individual active regions remain relatively constant.

Third, it provides new constraints for theoretical models of stellar dynamos, which must now account for this dichotomy between global and local magnetic properties.

\subsection{Implications for Stellar Age Determination}

The strong anticorrelation between spot coverage and stellar age identified in \cite{Araujo2025} has practical applications for stellar astrophysics. Traditional age-dating methods using isochrones can be significantly affected by starspots, which alter the apparent temperature and luminosity of the star. By accounting for the expected spot coverage based on other age indicators (such as rotation or activity), more accurate age estimates can be derived, particularly for young, active stars where the effect is most pronounced.

Our findings also strengthen the physical basis for gyrochronology and activity-age relations by clarifying the underlying mechanism driving the observed correlations. The decrease in spot coverage with age provides a direct link between the evolution of the stellar dynamo and observable changes in rotation and activity.

\section{Final Remarks}

This study has addressed a fundamental question in stellar evolution: does magnetic activity decline with age because individual starspots become magnetically weaker, or because the number of spots decreases? Our analysis provides a clear answer: it is the quantity, not the quality, that changes over stellar lifetimes. A star's magnetic activity decreases primarily because it produces fewer spots, not because its spots become magnetically weaker. The intrinsic \textit{quality} (magnetic strength) of the spots remains high and constant, while the \textit{quantity} (total area covered) is what truly changes over time.

Our key findings can be summarized as follows:
\begin{enumerate}
\item Starspot magnetic field strengths remain consistently high (2,700-4,600 G), suggesting a fundamental magnetic saturation plateau.
\item Spot coverage shows a strong anticorrelation with stellar age, decreasing from $\sim$29\% in young stars to $\sim$4\% in older stars.
\item The decline in stellar magnetic activity over time is primarily driven by a reduction in the total area of spots.
\end{enumerate}

These results have significant implications for stellar astrophysics:

\begin{itemize}
\item Improving Age-Dating: Accounting for spot coverage in stellar evolution models (isochrones) may lead to more precise age estimates, especially for young, active stars whose radii and temperatures are affected by high spot coverages \citep{Somers2015}.
\item Refining Dynamo Models: Models of how stars generate magnetic fields should focus on the mechanisms that control the rate of magnetic flux emergence to the surface, rather than the strength of individual flux tubes.
\end{itemize}


Future work will focus on expanding our sample to cover a wider range of stellar types and ages, further refining the relationship between spot properties and stellar parameters. Additionally, we plan to investigate potential star-planet magnetic interactions, examining whether planetary characteristics correlate with the properties of starspots, particularly those near the planet's transit path. This could provide insights into whether planets can modulate their host star's magnetic activity, a question of growing interest in the field of star-planet interactions.

\end{document}